\documentclass{aastex62}
\usepackage{float}

\begin{document}

\title{Discovery of a Weak CN Spectral Absorption Feature in Red Supergiant Stars in the Andromeda (M31) and Triangulum (M33) Galaxies}

\author[0000-0001-8867-4234]{Puragra Guhathakurta}
\affiliation{Department of Astronomy \& Astrophysics, University of California Santa Cruz, 1156 High Street, Santa Cruz, CA 95064, USA}

\author[0000-0001-8275-9181]{Douglas Grion Filho}
\affiliation{Department of Astronomy \& Astrophysics, University of California Santa Cruz, 1156 High Street, Santa Cruz, CA 95064, USA}

\author[0000-0002-4342-4626]{Antara R.\ Bhattacharya}
\affiliation{Harvard College, Cambridge, MA 02138, USA}

\author[0000-0001-8536-0547]{Lara R.\ Cullinane}
\affiliation{Leibniz-Institut f{\"u}r Astrophysik (AIP), An der Sternwarte
16, D-14482 Potsdam, Germany}
\affiliation{The William H.\ Miller III Department of Physics \& Astronomy,
Bloomberg Center for Physics and Astronomy, Johns Hopkins University, 3400 N.\ Charles Street, Baltimore, MD 21218, USA}

\author{Julianne J.\ Dalcanton}
\affiliation{Center for Computational Astrophysics, Flatiron Institute, 162 Fifth Avenue, New York, NY 10010, USA}
\affiliation{Department of Astronomy, University of Washington, Box 351580, Seattle, WA 98195-1580, USA}

\author[0000-0003-0394-8377]{Karoline M.\ Gilbert}
\affiliation{Space Telescope Science Institute, 3700 San Martin Drive, Baltimore, MD 2128, USA}
\affiliation{The William H.\ Miller III Department of Physics \& Astronomy,
Bloomberg Center for Physics and Astronomy, Johns Hopkins University, 3400 N.\ Charles Street, Baltimore, MD 21218, USA}

\author{Leo Girardi}
\affiliation{Osservatorio Astronomico di Padova -- INAF, Padova 35136, Italy}

\author{Anika Kamath}
\affiliation{Stanford University, 450 Jane Stanford Way, Stanford, CA 94305, USA}

\author[0000-0001-6196-5162]{Evan N.\ Kirby}
\affiliation{Department of Physics \& Astronomy, University of Notre Dame, 225 Nieuwland Science Hall, Notre Dame, IN 46556, USA}

\author{Arya Maheshwari}
\affiliation{Princeton University, Princeton, NJ 08540, USA}

\author[0000-0002-9137-0773]{Paola Marigo}
\affiliation{Osservatorio Astronomico di Padova -- INAF, Padova 35136, Italy}

\author[0000-0002-3361-2893]{Alexandra Masegian}
\affiliation{Department of Astronomy, Columbia University, New York, NY 10027, USA} 

\author[0000-0001-8481-2660]{Amanda C.\ N.\ Quirk}
\affiliation{Department of Astronomy, Columbia University, New York, NY 10027, USA}

\author{Rachel Raikar}
\affiliation{Department of Earth \& Planetary Sciences, University of California Santa Cruz, 1156 High Street, Santa Cruz, CA 95064, USA}

\author[0009-0003-5497-3932]{Stanley M.\ Rinehart V}
\affiliation{Department of Astronomy \& Astrophysics, University of California Santa Cruz, 1156 High Street, Santa Cruz, CA 95064, USA} 


\author{Caelum J.\ Rodriguez}
\affiliation{Department of Astronomy \& Astrophysics, University of California Santa Cruz, 1156 High Street, Santa Cruz, CA 95064, USA}


\author[0000-0002-7502-0597]{Benjamin F.\ Williams}
\affiliation{Department of Astronomy, University of Washington, Box 351580, Seattle, WA 98195-1580, USA}

\begin{abstract}

\noindent
Using Keck DEIMOS spectra of stars in the Andromeda (M31) and Triangulum (M33) galaxies, selected from the large multi-band (near ultraviolet, visible light, and near infrared) {\it Hubble Space Telescope\/} surveys PHAT and PHATTER, respectively, we have identified a subset of stars that contain a previously unnoticed weak spectral absorption feature around 8000\,\AA\ (0.8\,$\mu$m). This absorption feature appears to be associated with the cyanogen (CN) molecule. Strong CN spectral absorption is a standard feature of carbon stars, which are thought to be intermediate mass (2--3~$M_\odot$) stars with $\rm C/O>1$ in the thermally-pulsating asymptotic giant branch phase of stellar evolution. However, the stars that are the focus of this paper are characterized by a weak version of this CN spectral absorption feature in a spectrum that is otherwise dominated by normal O-rich spectral absorption lines such as TiO and/or the Ca near infrared triplet. We have dubbed these stars ``weak CN'' stars. We present an automated method for identifying weak CN stars in M31 and M33, and examine their photometric properties in relation to model isochrones and stellar tracks. We find that weak CN stars tend to be fairly localized in color-magnitude space, and appear to be red supergiant stars with masses ranging from 5--10~$M_\odot$, overall lifetimes of about 40--50~Myr, and currently in the core He burning phase of stellar evolution.

\end{abstract}

\section{Introduction}\label{sec:intro}

Near field cosmology and galactic archaeology involve careful analyses of the fossil record of galaxy formation. There have been several large surveys of the resolved stellar population of the Milky Way (MW) over the last few decades, and the remarkable {\it Gaia\/} mission has ushered in the so-called Galactic Renaissance \citep[e.g.,][]{Eggen62, Searle78,Yanny00, Yanny09, Newberg02, Belokurov06, Helmi18, Mor19, Conroy19, Cunningham19a, Cunningham19b, Feuillet20, Naidu20, Rockosi22, McKinnon23}. The Local Group galaxies Andromeda and Triangulum (M31 and M33, respectively) are also excellent testbeds for the study of galaxy formation and evolution due to a combination of two factors: their relative proximity allows us to observe them in detail, while our external vantage point provides a global perspective. Several photometric and spectroscopic surveys of the resolved stellar populations of M31 and M33 have examined the history of their satellite mergers/halo accretion, disk dynamical heating, star formation, and chemical enrichment \citep[e.g.,][]{Ibata01,Ibata07,Ibata14, Brown03, Brown06a, Brown06b, Irwin05, Guhathakurta05,Guhathakurta06,Blitz06, Gilbert06,Gilbert20,Gilbert22, Chapman06, McConnachie06,Dalcanton12, Dorman12,Dorman13,Dorman15, Williams15b, Williams17, Williams21, Lewis15, Quirk19, Quirk22, Escala20, Kirby20, Lazzarini22, Johnson22, Smercina23}.


In addition to using stars to study galaxy formation and evolution, one can use nearby galaxies as laboratories for the study of stellar evolution. Each nearby galaxy presents an easily accessible sample of stars at a common distance, which removes an important confounding factor in the characterization of mixed stellar populations. In particular, large spectroscopic samples of M31 and M33 stars, originally obtained for the purpose of studying galaxy formation and evolution, can also be used to study rare phases of stellar evolution.
    
Carbon (C) stars represent one such example of a rare phase of stellar evolution among intermediate mass stars. Due to their easily identifiable spectra, luminous C stars have been classified and characterized in substantial numbers in M31 and M33 \citep{Battinelli04a,Battinelli04b,Boyer13, Hamren15, Hamren16, Girardi20}. In addition to providing insight into a rare phase of stellar evolution (see below), C stars have also been used to map the morphological and kinematical structure, star formation history, and metallicity of their host galaxies \citep[e.g.,][]{Battinelli05, Rowe05, Demers07, Cioni08,Javadi13,Javadi17, Rezaeikh14, Huxor15,Hamedani17,Hasemi19, Quirk22, Gilbert22}. These stars have also been used as standard candles to improve the calibration of the extragalactic distance scale \citep{Pierce00, Huang18, Huang20,Madore20, Ripoche20}.

Most luminous C stars are intermediate age ($\sim$ few Gyr) asymptotic giant branch (AGB) stars with a C/O ratio greater than 1. The excess C found in most C stars is commonly thought to have been built up through the standard stellar evolutionary framework, in which C is built up in the outer layers via the third dredge-up (TDU) process in thermally pulsating AGB (TP-AGB) stars. This stage is characterized by unstable double shell burning. The TDU occurs when the He-burning shell around the AGB star's inert C+O core ignites, extinguishing the outer H-burning shell. This allows for the outer convective envelope of the star to reach the region within the shell and pull (or dredge) C up to the surface. Over time and through successive dredge-ups, the C/O ratio increases, until eventually the number of free C atoms exceeds the number of free O atoms in the star's atmosphere, nominally transforming it into a C star.
A second and rarer channel for C star formation applies to stars whose masses are lower than what is necessary for TDU, such as CH stars, dwarf C (dC) stars, and C-enchanced metal-poor (CEMP) stars. CEMP stars are especially important for investigating the origins of the very first Population~III stars \citep[e.g.,][]{Meynet06,Jeon21}.

At higher metallicities, C can be generated by faint ``mixing and fallback'' supernovae \citep{Umeda05}. The products of either type of supernovae (Ia and core collapse) can enrich the next generation of stars, some of which are low-mass and still observable today. An alternative scenario is ``extrinsic'' enrichment, where C is transferred to a star after it is born. The most common donor for extrinsic enrichment is a C-rich AGB companion \citep{deKool95, Frantsman97, Izzard04}.

While there have been studies of dC stars in the MW \citep[e.g.,][]{Plant16}, these dC stars are far too faint to be observable at the distance of M31 and M33 given the detection threshold of current spectroscopic surveys. Therefore, when we refer to C stars later in this work, we refer to only the TP-AGB variety.

As part of the SPLASH and TREX surveys of the resolved stellar population of the disks of M31 and M33, respectively, we have obtained tens of thousands of optical stellar spectra using the DEIMOS spectrograph \citep{Faber03} on the Keck~II 10-m telescope \citep{Dorman12, Dorman13, Dorman15, Gilbert22, Quirk22, Cullinane23}. During our visual inspection of these spectra, it was relatively easy to identify TP-AGB C stars via (amongst other things) their characteristic strong `W'-shaped cyanogen (CN) feature at 8000\,\AA. We also found a set of stars whose spectra contain a subtle version of the 8000\,\AA\ CN band with the rest of the spectrum resembling that of an O-rich star (with TiO bands and/or the Ca II infrared triplet, for example). We have named these stars ``weak CN'' stars, given that their CN absorption feature is much weaker than that found in C stars. In \S\,3.1, we estimate that the CN features in weak CN stars are on average $\sim 6$ times weaker when compared to C stars.

This paper reports on the discovery of these weak CN stars in M31 and M33 based on Keck DEIMOS spectra from the SPLASH and TREX surveys, respectively. The underlying astrophysical explanation of the weak CN phenomenon remains unknown to us at this time, but we report on our finding nevertheless. In the same spirit, in a companion paper \citep[][hereafter Paper~II]{GrionFilho25}, we present the discovery and characterization of weak CN stars in the Large Magellanic Cloud (LMC) using the \citet{Olsen11} sample of spectra obtained with the Hydra multi fiber spectrograph on the Cerro Tololo Inter-American Observatory 4-m Blanco telescope \citep{Bardem&Ingerson98}. Paper~II expands upon the findings presented in this paper and compares the weak CN population across three host galaxies: M31, M33, and the LMC, a sequence of decreasing galaxy stellar mass, decreasing mean metallicity, and increasing specific star-formation rate. Paper~II measures the age of weak CN stars in these three galactic hosts using a Bayesian approach to isochrone fitting and compares them to low-resolution atmospheric stellar models in an attempt to understand their origin. As we show in this paper (\S\,4 onwards), the weak CN phenomenon appears to be characteristic of young supergiants in a specific region of the color-magnitude diagram (CMD). For starters, studying the properties of weak CN stars in different galactic environments will hopefully shed light on how dependent this phenomenon is on the properties of the host galaxy. Ultimately, a deeper understanding of the astrophysical cause of the unusual surface chemistry of weak CN stars and the red supergiant phase of stellar evolution they appear to be associated with (see \S\,4.1) could provide insights into internal processes in massive evolved stars, the chemistry and energetics of their surrounding interstellar medium, and the chemical evolution of galaxies \citep{Wood83, Massey05, deBeck10, Boyer12, Boyer24}.

Next, we outline the main sections of this M31/M33 weak CN paper. In \S\,\ref{sec:data}, we describe the M31 and M33 resolved stellar population photometric and spectroscopic datasets. In \S\,\ref{sec:auto_class}, we present the discovery of weak CN stars, compare the weak CN spectrum to a C star spectrum, and describe a new automated spectral classification method for weak CN stars. In \S\,\ref{sec:rsg}, we obtain rough mass and age/lifetime estimates for weak CN and C stars by comparing their position in various CMDs to theoretical isochrones and stellar tracks. In \S\,\ref{sec:disc}, we discuss the similarities and differences between weak CN stars in M31 and M33, and briefly explore the prospect of studying such stars in the MW and other galaxies. Finally, in \S\,\ref{sec:conc}, we list the conclusions of this work and possible future directions.

We adopt the following parameters in this paper: true distance moduli of $(m-M)_0=24.4$ and 24.6~mag for M31 and M33, respectively \citep{Holland98,Breuval23}, and mean total extinction (foreground MW plus M31/M33) of $A_V^{\rm tot}=1.0$ and 0.4~mag for M31 and M33, respectively \citep{Wang22a,Wang22b}.

\section{Data}\label{sec:data}

This study relies on the synthesis of multiband photometric data and spectra. Moreover, the photometric data were an essential element of the design of the spectroscopic survey. As described below, these two kinds of data were obtained using a few different combinations of telescopes, instruments, and instrumental configurations.

\subsection{Hubble Space Telescope Photometry} \label{sec:hst_phot}
Brightness and color measurements of weak CN stars, C stars, and other stars in M31 and M33 are derived from {\it Hubble Space Telescope\/} ({\it HST\/}) imaging in five filters: F336W in the near ultraviolet (similar to ground based $U$) using Wide Field Camera 3 ultraviolet/visible light channel (WFC3/UVIS), F475W and F814W in the optical (similar to ground based $B$ and $I$, respectively) using Advanced Camera for Surveys wide field channel (ACS/WFC), and F110W and F160W in the near infrared (similar to ground based $J$ and $H$, respectively) using WFC3 near-infrared channel (WFC3/IR).

The M31 {\it HST\/} data are from the Panchromatic Hubble Andromeda Treasury (PHAT) survey, an 828-orbit multi-cycle treasury program that imaged the resolved stellar population in the northeastern portion of the galaxy's disk \citep{Dalcanton12}. The PHAT survey footprint is in the form of 23 ``bricks'' covering 0.5-deg$^2$ in a tiling pattern. The M33 {\it HST\/} data are from the Panchromatic Hubble Andromeda Treasury: Triangulum Extended Region (PHATTER) survey, a 108-orbit program that imaged the resolved stellar population in the central high surface brightness portion of the galaxy's disk \citep{Williams21}. The PHATTER survey footprint is in the form of 3 contiguous bricks covering a roughly $12' \times 18'$ rectangular region.

Each PHAT/PHATTER brick has dimensions of $6' \times 12'$ and consists of a $3 \times 6$ tiling pattern of fields imaged with WFC3/IR, WFC3/UVIS, and ACS/WFC. Of the three imaging cameras/channels used in the PHAT/PHATTER surveys, WFC3/IR has the smallest field of view and sets the spacing between adjacent fields in the tiling pattern that makes up each brick. The amount of overlap between adjacent fields in a brick is smallest for WFC/IR and largest for ACS/WFC.

It should be noted that the PHAT survey of M31 and PHATTER survey of M33 are based on six-filter photometry: the ultraviolet F275W filter using WFC3/UVIS in addition to the five filters mentioned above. However, the photometry in the F275W band is too shallow for it to be useful in this work.

The photometric analysis in this paper (\S\,\ref{sec:cmd}) is based on the PHAT and PHATTER brick-wide source catalogs derived from forced PSF-fitting. \citet{Dalcanton12} and \citet{Williams14, Williams21, Williams23} provide detailed descriptions of the PHAT and PHATTER survey/tiling strategy and six-filter photometry procedure. In total, PHAT has obtained photometric measurements for over 117 million stars in M31, while PHATTER has obtained measurements for around 22 million stars in M33. The final PHAT and PHATTER photometry catalogs are available via the Mikulski Archive for Space Telescopes \citep[MAST ---][]{PHAT_MAST,PHATTER_MAST}.

\subsection{Keck DEIMOS Spectroscopy}\label{sec:data spec}

\subsubsection{Target Selection and Survey Design} \label{sec:targ_sel_srvy_design}
Spectroscopy was carried out using DEep Imaging, Multi-Object Spectrograph \citep[DEIMOS;][]{Faber03} multislit masks on the Keck~II 10-m telescope located on the summit of Maunakea, Hawaii. Spectroscopic targets were selected from regions of the sky with {\it HST\/} imaging---specifically, the PHAT survey region of M31 and the PHATTER survey region of M33---in order to ensure a consistent set of photometric bandpasses across the M31 and M33 samples. Two DEIMOS instrument settings were used: (1)~medium resolution ($R\sim5000$) using the 1200G (1200 line mm$^{-1}$) grating centered on 7800~\AA\ and covering the approximate wavelength range 6500--9000~\AA\ for masks targeting red stars such as red giant branch (RGB) stars, AGB stars, red supergiants, etc.; and (2)~low resolution ($R\sim2000$) using the 600ZD (600 line mm$^{-1}$) grating centered on 7200~\AA\ and covering the approximate wavelength range 4700--9700~\AA\ for masks targeting a wider range of stellar types such as red stars, blue main sequence stars, blue/yellow supergiants, etc.).

In an effort to avoid sources that are likely to be blended in ground based seeing, we adopted the isolation criterion that was used by \citet{Dorman12} in the SPLASH survey of M31's northeastern disk. The criterion excludes stars that have neighbors that are too close and/or too bright that would contaminate the target star's light in ground based seeing. Our spectroscopic survey avoided the crowded central bulge of M31 altogether and the isolation criterion was applied to the rest of the PHAT sample and all of the PHATTER sample; the criterion was especially useful in the crowded central regions of M33 in the PHATTER sample.
%

The M31 data set consists of 20 medium resolution masks and 21 low resolution masks. The M31 data were obtained in the course of the ongoing Spectroscopic and Photometric Landscape of Andromeda’s Stellar Halo (SPLASH) survey \citep{Guhathakurta05, Guhathakurta06, Gilbert06}, specifically the portion of the survey that covered the PHAT survey region in the northeastern portion of M31's disk during the 2010B, 2011B, and 2012B observing seasons \citep{Dorman12, Dorman13, Dorman15}. The M33 data set consists of 22 low resolution masks. The M33 data were obtained in the course of the ongoing TRiangulum EXtended (TREX) survey \citep{Gilbert22, Quirk22, Cullinane23}, specifically the portion of the survey that covered the PHATTER survey region in the central portion of M33's disk during the 2018B, 2019B, and 2021B observing seasons.

It is important to note that there are two differences between the M31 and M33 datasets: (1) the M31 dataset includes fainter (lower luminosity) stars than the M33 dataset because of differences in spectroscopic target selection strategy between the SPLASH and TREX surveys; and (2) M31 stars span a wider range of colors than M33 stars because M31 has a higher mean metallicity/metallicity spread and higher/more variable dust reddening. We explore the effects of these two differences, one observational and one astrophysical, on our weak CN star analysis in \S\S\,\ref{sec:omega} and \ref{sec:m31m33comp}.

There is substantial overlap across the footprints of the TREX DEIMOS masks in the PHATTER region of M33, and this has allowed for $\sim400$ instances of repeated spectroscopic measurements (i.e., instances of the same star being included as a target in more than one DEIMOS mask). There is not nearly as much overlap across the SPLASH DEIMOS masks in the PHAT region of M31, and there are only a few tens of repeats. All of these spectra, including the repeats, are analyzed in this paper.


\subsubsection{Data Reduction and Visual Inspection} \label{sec:spec_data_red_zspec}
The raw data from these DEIMOS multislit masks were processed using the IDL-based {\tt spec2d} data reduction pipeline \citep{Newman13}. The output of this pipeline is a 2D spectrum and a 1D spectrum for each spectroscopic target on the mask. The 1D spectra were then run through cross-correlation analysis using the IDL-based {\tt spec1d} software pipeline. The reader is referred to \citet{Guhathakurta06} for the details of these data reduction steps.

The 2D spectrum of each slit on each mask and the inverse variance (ivar) weighted Gaussian smoothed 1D spectrum of the corresponding target were visually inspected using the IDL-based {\tt zspec} graphical user interface. The purpose of this visual vetting is three-fold: (1) to identify data reduction issues ranging from severe issues that lead to a catastrophic failure to measure the redshift ({\tt zqual} flag set to $-$2) to mild issues associated with imperfect wavelength calibration at the short wavelength end of the spectrum, bad extraction window used to extract the 1D spectrum from the 2D spectrum, apparent continuum discontinuity across DEIMOS's blue and red CCDs, bad columns, CCD edges, bad sky subtraction, vignetting, scattered light, etc.; (2) to assess the quality of the redshift measurement yielded by the {\tt spec1d} pipeline: secure (${\tt zqual}=4$), marginal (${\tt zqual}=3$), unreliable (${\tt zqual}=2$), or superseded by a visual measurement (${\tt zqual}=1$) --- see \citet{Guhathakurta06}; and (3) identification of astrophysically interesting sources such as unusual stars (e.g., C and weak CN stars, foreground stars), background sources (e.g., distant galaxies and quasars), emission lines from the diffuse interstellar medium of M31/M33, serendipitously detected neighbors, etc. 

\subsubsection{Identification and Removal of Foreground Milky Way Stars}\label{sec:fg_mw_stars}


Our analysis focuses on the disks of M31 and M33. Foreground contamination is low throughout, but more significant in the M31 sample given that it consists of the reshifted north-eastern half of the galaxy. 

\citet[][see \S\,3.1.1 of their paper]{Dorman15} removed foreground MW stars from the PHAT-selected Keck DEIMOS spectroscopic sample in the region of M31's northeastern disk using techniques developed by \citet{Gilbert06}. Foreground MW M dwarfs are easily identifiable via the presence of a strong Na~I doublet at $\sim8200$~\AA. Foreground MW M dwarf stars were visually identified and removed from the PHATTER-selected Keck DEIMOS spectroscopic sample in the region of M33's disk. Warmer foreground MW star contaminants (e.g., early K dwarfs) would not have been identified in this visual identification, but such stars are likely to constitute and even smaller fraction of the overall M33 sample. Finally, foreground MW dwarf C stars are likely to be a negligible source of contamination, because the dwarf C phenomenon is known to be much rarer than the AGB C star (and weak CN star) phenomenon.

Similar methods were used to identify and remove foreground MW stars in the M33 sample. These MW contaminants constitute about 2\% of the overall sample, and their heliocentric velocities have a mean and standard deviation of $-33.3$ and 56.2~km~s$^{-1}$, respectively. The velocity distribution of these foreground MW stars is offset from the main velocity distribution of M33 disk stars, whose mean heliocentric velocity is $-196$~km~s$^{-1}$. 

In order to identify additional foreground MW star contaminants in our Keck DEIMOS M31/M33 spectroscopic samples, we cross-matched the samples with the \textit{Gaia} DR3 catalog \citep{Gaia16, Gaia23}, using a $0\farcs1$ matching radius. We removed an additional 2 (18) stars from the M31 PHAT (M33 PHATTER) samples whose \textit{Gaia} parallax and proper motion values are consistent with them being MW stars. However, the majority of the stars in our spectroscopic samples are faint and located in crowded fields which makes their \textit{Gaia} astrometry too imprecise to reliably distinguish between M31/M33 stars and foreground MW contaminants.

\subsubsection{Processing of 1D Spectra} \label{sec:proc_1dspec}
The 1D spectrum of each M31/M33 member star with a secure or marginal redshift measurement is Doppler shifted to the rest frame using the observed (not heliocentric) velocity of the star. Each Doppler shifted spectrum is then rebinned to a common wavelength grid. This step is carried out for both 1200G (medium resolution) and 600ZD (low resolution) spectra.

The flux and inverse variance of each spectrum are then normalized as follows:

\begin{equation}
    F = F_{\rm raw} / \langle{F}\rangle
\end{equation}

\begin{equation}
    {\rm ivar} = \left(\frac{\sigma_{\rm raw}}{\langle{F}\rangle} \right)^{-2} = {\rm ivar_{raw}} \langle{F}\rangle^2
\end{equation}

\noindent
where $F_{\rm raw}$ and $F$ are the raw and normalized fluxes, respectively, $\rm ivar_{raw}$ and $\rm ivar$ are the raw and normalized inverse variances, respectively, and $\langle{F}\rangle$ is median flux of the raw spectrum over the wavelength range 7800--8200~\AA.\\

\vskip 0.5truecm

In summary, we limit our analysis to the intersection of the PHAT photometric sample and the SPLASH spectroscopic sample which covers the northeastern half of M31's disk \citep{Dalcanton12, Williams14, Dorman15} and to the intersection of the PHATTER photometric sample and TREX spectroscopic sample which covers most of the high surface portion of the disk of M33 \citep{Williams21, Gilbert22, Quirk22, Cullinane23}. We limit our analysis to M31/M33 member stars with a secure or marginal redshift measurement, selecting only those with a {\tt zqual} flag equal to 1, 3, or 4 \citep{Dorman15}. This yields a final spectroscopic sample of 8009 stars in M31 and 2665 stars in M33. 

\section{Identification and Classification of Weak CN and Carbon Stars}\label{sec:auto_class}

\subsection{The Strength of the ``Weak CN" Spectral Feature}\label{sec:spec}

The optical spectra of C stars are distinguished by prominent CN features at 7700-8200~\AA\ and C$_2$ bands at 6100--6600~\AA. Previous authors have used the CN band to identify C stars, either by using narrowband filters centered on CN and its O-rich counterpart TiO \citep{Nowotny03, Battinelli04a, Battinelli04b, Wing07} or by cross-correlating spectra against templates with and without CN \citep{Hamren15, Hamren16}. This spectral feature is not generally present in normal O-rich stars (M giants).
 
\begin{figure}[H]
    \centering
    \includegraphics[width=0.99\textwidth]{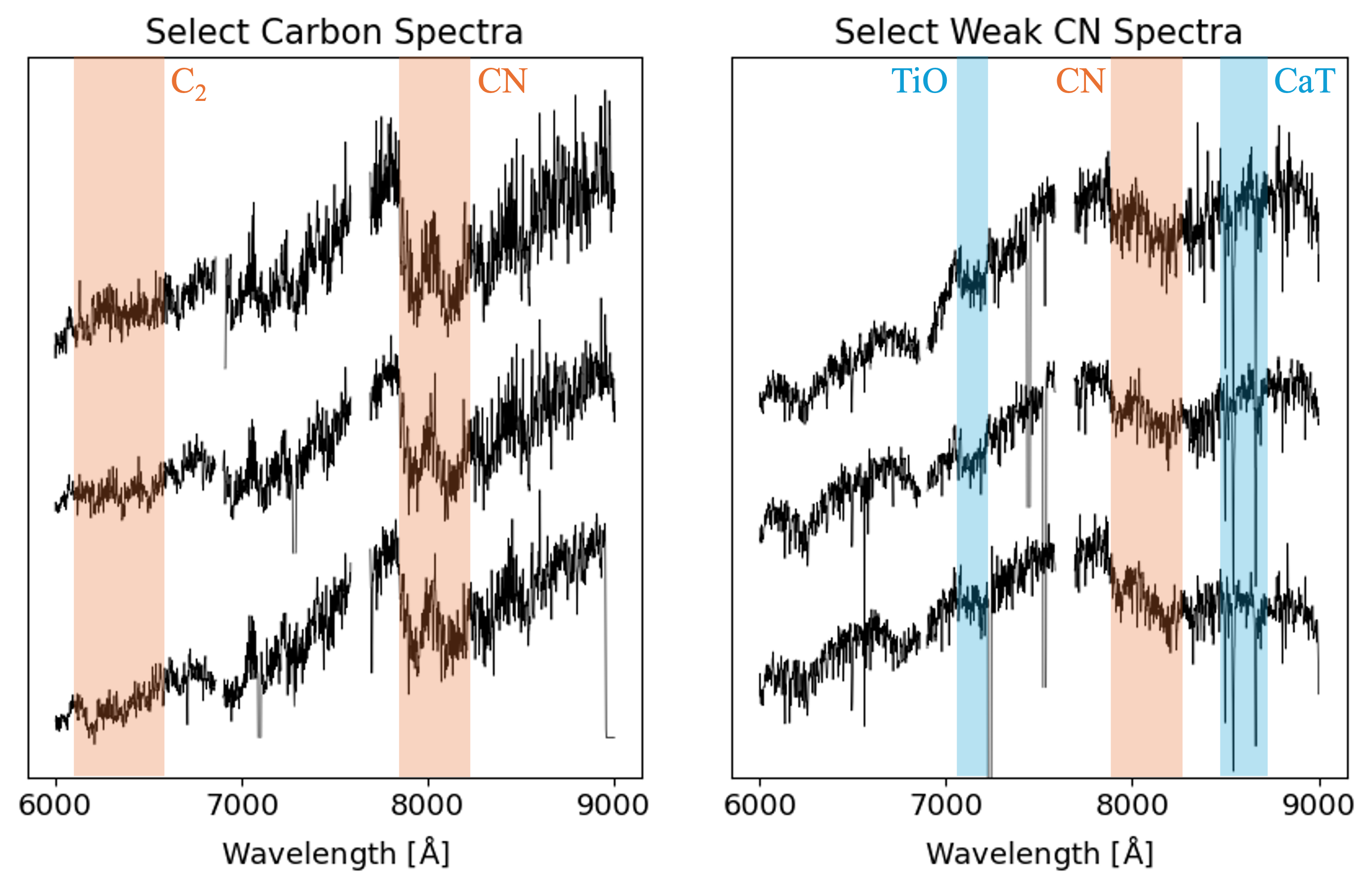}
    \caption{A few examples of Keck DEIMOS 600ZD spectra of C and weak CN stars in M31 (left and right panels, respectively). The spectra are not flux calibrated; the y axis is relative instrumental flux with an arbitrary offset for the sake of plotting clarity. The 7700--8300~\AA\ region highlighted in pink in both panels contains the ``W"-shaped CN absorption feature, while the 6000--6500~\AA\ pink band in the left panel contains C$_2$ absorption features. The 7000--7100~\AA\ and 8500--8700~\AA\ regions highlighted in blue for weak CN stars (right panel) contain TiO bands and the Ca~II triplet, respectively; these spectral absorption features are typically associated with O-rich stars. The 7700--8300~\AA\ CN feature is significantly weaker in weak CN stars (right panel) than in C stars (left panel). Gaps in the spectra at $\approx6850$~\AA\ and $\approx7600$~\AA\ correspond to the strong B and A telluric absorption bands, respectively, and have been masked out. Each spectrum contains an instrumental artifact, a sharp rectangular dip in the 7000--7500~\AA\ range caused by the gap between the blue and red CCDs in DEIMOS.}
    \label{fig:spectracomp1}
\end{figure}

\begin{figure}[H]
    \centering
    \includegraphics[width=0.99\textwidth]{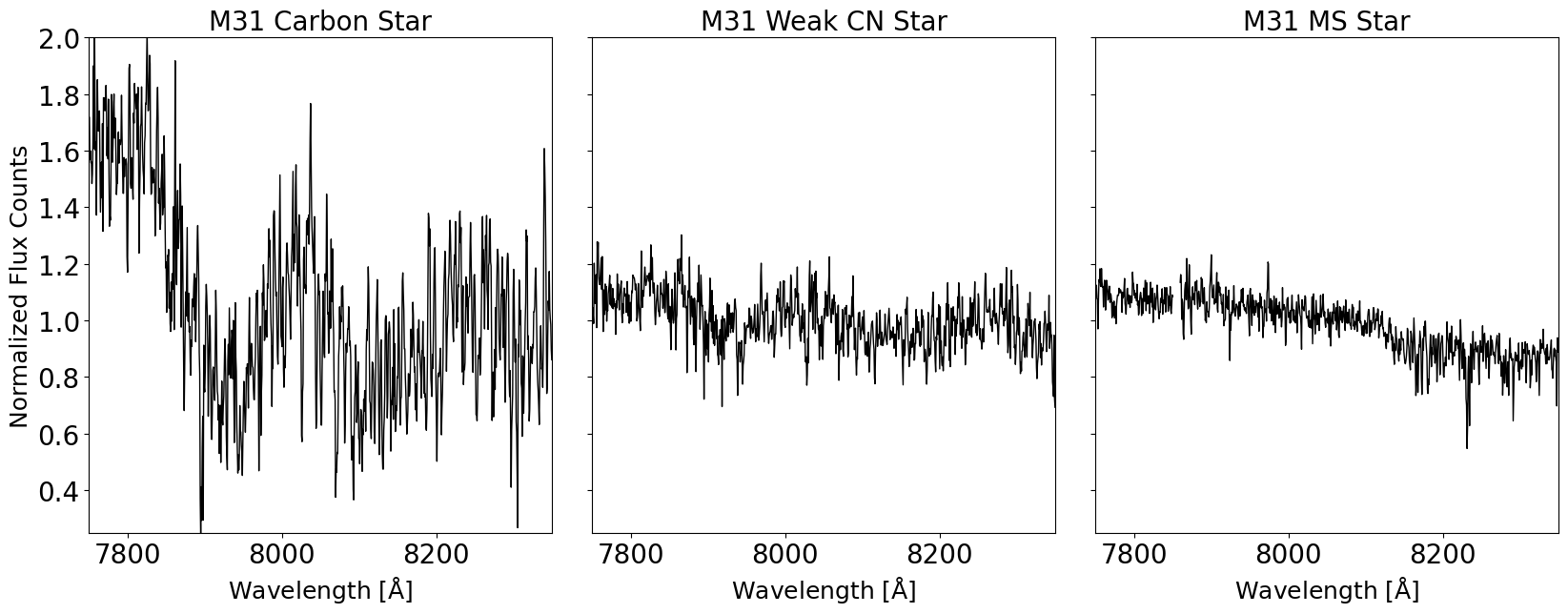}
    \caption{Comparison of normalized spectra of arbitrarily chosen examples of a C, weak CN, and massive main sequence star (left, center, and right, respectively). The 7800--8200~\AA\ ``W"-shaped feature of the carbonaceous molecule CN is strong in the C star spectrum. Although weak, the feature is still distinguishable in the weak CN spectrum, but absent in the main sequence star spectrum}.
    \label{fig:spectracomp2}
\end{figure}

As mentioned in \S\,\ref{sec:intro}, our visual inspection of stellar spectra in M31 led to a discovery of a group of stars that contain weak absorption features associated with the carbonaceous molecule cyanogen (CN) along with absorption features associated with O-rich stars, such as titanium oxide (TiO) bands and the near infrared calcium triplet\citep{Kamath16, Chauhan21,Rodriguez21, Bhattacharya23}. A few examples of C stars and weak CN stars in M31 are shown in Figure~\ref{fig:spectracomp1}. Later, during our visual inspection of M33 spectra, we noticed stars with similar spectral characteristics. These mysterious stars are tentatively dubbed ``weak CN'' stars, in reference to the fact that the ``W''-shaped CN spectral absorption feature is much weaker in these stars than in C stars---e.g.,~see Figures~\ref{fig:spectracomp1} and \ref{fig:spectracomp2}. In fact, CN absorption is the most prominent optical spectral feature of C stars, but is entirely absent from the spectra `normal' M stars that contain O-dominant atmospheres. We had therefore initially hypothesized that the weak CN subpopulation may be associated with the C star subpopulation, perhaps representing an intermediate phase of stellar evolution on the path to a star developing a C-rich atmosphere and becoming a C star.


\begin{figure}[H]
    \centering
    \includegraphics[width=0.99\textwidth]{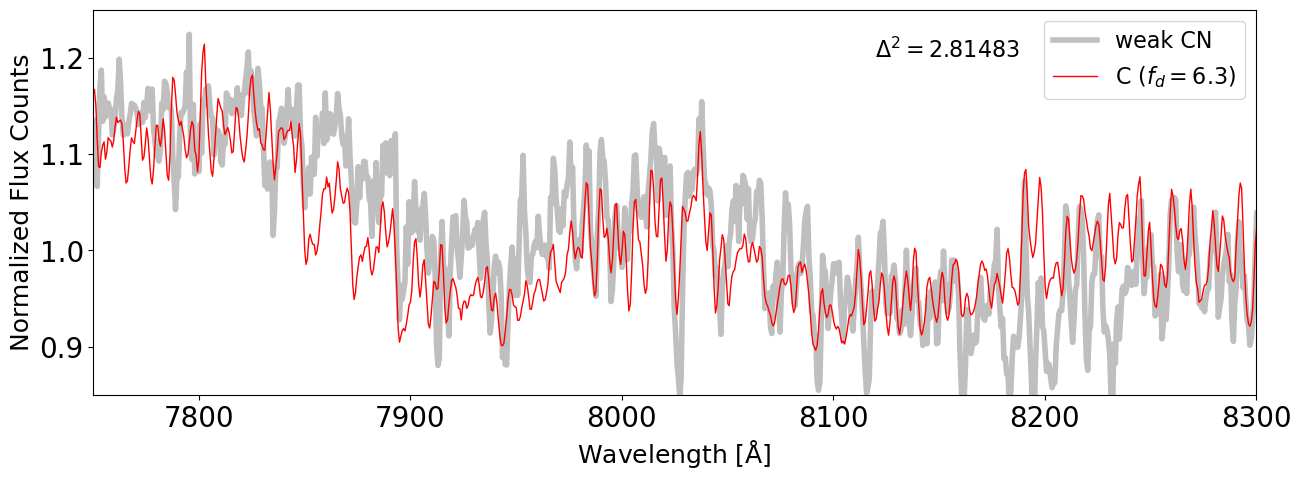}
    \caption{Zoomed-in comparison between the weak CN coadded spectrum (bold gray line) and the diluted C star coadded spectrum (thin red line). We find a best dilution factor of $f_d=6.3$ and note that there are some features that are stronger in the weak CN than the diluted C star spectra and vice versa, but overall the two match well. Our difference factor $\Delta^2$ is printed near the top right.  
    In other words, cyanogen absorption appears to be about six times weaker on average in weak CN stars than in C stars.} 
    \label{fig:dilution}
\end{figure}

In order to compare the strength of the CN spectral absorption feature between weak CN and C stars, we rely on high signal-to-noise coadded spectra of these two subpopulations (see \S\,\ref{sec:templates} for the details of the construction of these coadded spectral templates). For this comparison, we merely overlay the two coadded spectra on top of each other, but ``dilute'' the coadded C star spectrum by a factor $f_d$. By adding a constant ($f_d - 1$) to the C star spectrum and renormalizing the resulting spectrum by $f_d$, we effectively weaken the strength of all spectral absorption features by a factor $f_d$. This is a purely mathematical operation with no astrophysical basis, but it allows us to quantify how much weaker the CN spectral absorption feature is in weak CN stars relative to C stars (Figure \ref{fig:dilution}). From this cursory analysis, we note that the weak CN and diluted C star coadded spectra share many common absorption features across the CN wavelength range (7750--8300~\AA).

We make a rough measurement of how ``similar'' the diluted C star spectrum is to the weak CN spectrum by summing the squared difference of the two over the relevant wavelength range ($\Delta^2$) and minimizing this quantity to determine the best fit dilution factor $f_d$. We find that the CN absorption is around $6.3\times$ weaker on average in weak CN stars than in C stars, with a corresponding minimum squared difference of $\Delta_{\rm min}^2=2.8$ (Figure \ref{fig:dilution}).

Given the radial metallicity gradient in the disk of M31, we test the effect of metallicity on the strength of the weak CN spectral absorption feature by measuring $f_d$ across three radial bins in the disk of M31. We list the best fit dilution factor $f_d$ (corresponding $\Delta_{\rm min}^2$) in order of increasing radius, and find that it does not vary monotonically with radius: 7.0 (3.0), 6.0 (3.4), and 6.5 (3.3). The dilution factor of $f_d=6.3$ estimated above appears to be representative of the entire M31 weak CN star population.

\subsection{Empirical Weak CN and Carbon Star Spectral Templates}\label{sec:templates}
In order to investigate the properties of this exotic group of stars (and indeed decide whether they are a new group entirely or just a previously unseen feature of a well understood group), we devise a method for identifying all such weak CN stars within our sample. We begin by selecting a small initial sample of visually classified weak CN and C stars. From these spectra, we generate a high signal-to-noise coadded spectral template, which the rest of the dataset will be compared against. The number of visually selected stars used for this initial step is somewhat arbitrary, though of course more stars will improve the signal-to-noise ratio of the resulting template. For M31, we visually identified around 70 stars from each of the weak CN and C star populations to generate their respective templates. For M33, we identified 140 weak CN stars and 34 C stars. We generate a separate template for each galaxy to best differentiate between the weak CN and C star populations within each galaxy.

    
Coadded spectral templates are constructed by calculating the normalized inverse variance (ivar) weighted average of the normalized flux $F$ at each wavelength $\lambda$ (spectral pixel) as follows:

\begin{equation}
    F_{\rm coadd} (\lambda) = \frac{1}{\sum_i {\rm ivar}_i (\lambda)} \sum_i F_i(\lambda){\rm ivar}_i (\lambda)
\end{equation}

\noindent
where the index $i$ runs over the $N$ spectra that are being coadded.

We generate two sets of spectral templates for each group: one limited in wavelength to the range where the CN feature is present and another that is not restricted in wavelength. A star is excluded from both coadded templates and any other spectral analysis if $>10\%$ of the flux values in the CN wavelength range are invalid (NaN values). In M33, the number of stars removed due to this cut was 10. In M31, no stars had to be removed.

\subsection{Comparisons to Empirical Templates (COMET)}\label{sec:comet}

\begin{figure}[H]
    \centering
    \includegraphics[width=0.7\textwidth]{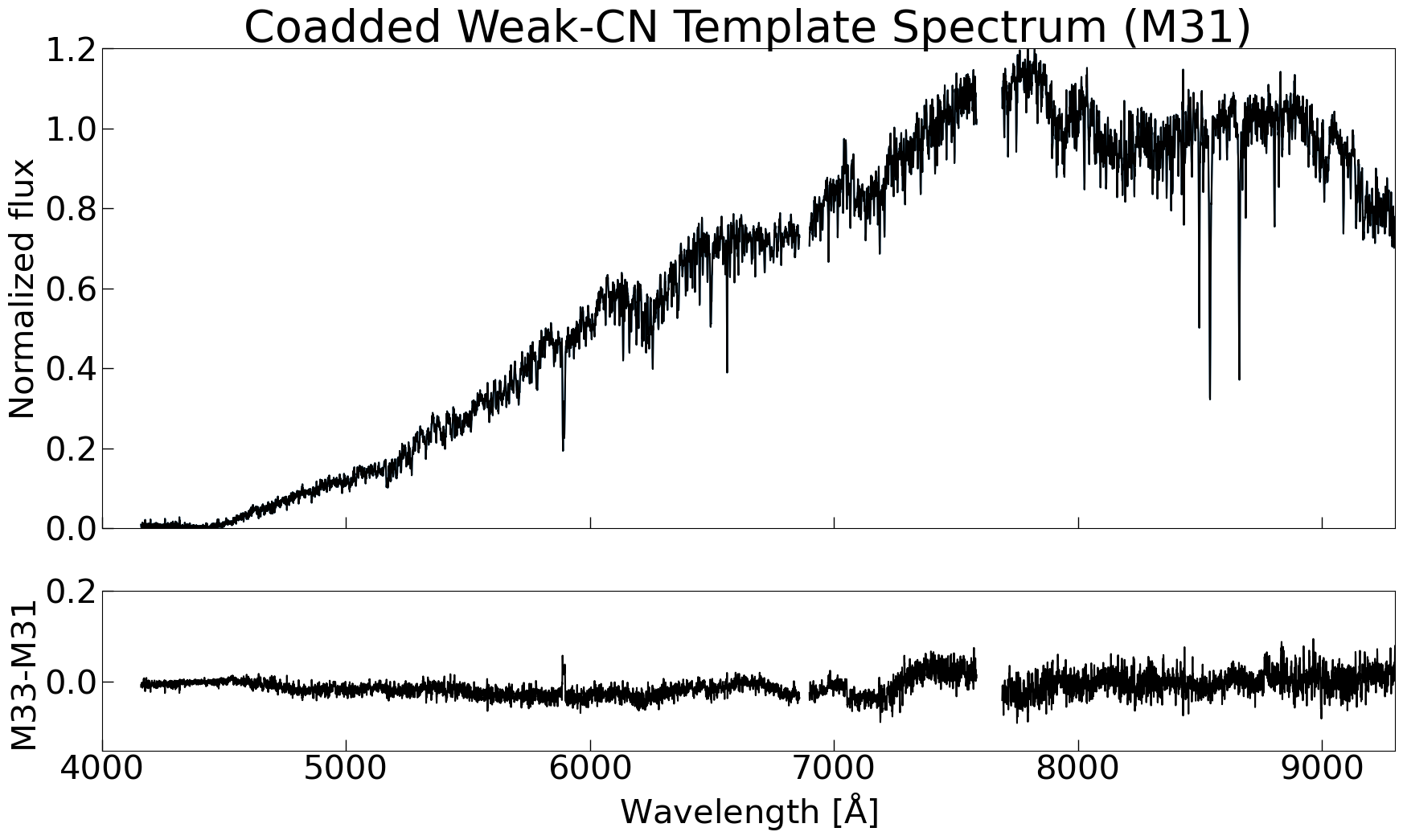}
    \caption{\textit{Top:} Template coadded spectrum of all visually classified weak CN stars in M31. Telluric absorption bands have been removed, leaving behind two gaps at roughly 6900 and 7700 \AA. \textit{Bottom:} Difference between the M31 and M33 weak CN spectral templates. The two spectral templates are essentially identical; some of the small differences observed could be noise artifacts while others may be the result of the difference in mean metallicity between M31 and M33.}
    \label{fig:template}
\end{figure}

With our spectral templates completed, we now discuss the method used for applying them in the classification of our sample and subsequent identification of weak CN stars. For the complete M31 weak CN template, see the top panel of Figure \ref{fig:template}. The bottom panel of the figure shows the difference between the M33 and M31 templates. The standard deviation of the difference is as low as 0.025, indicating that the two templates are essentially identical.

First, before any wavelength-by-wavelength flux comparison is made an individual spectrum and the coadded spectral template, the overall slope of the underlying spectral continuum of the star is matched to that of the coadded spectral template over the wavelength range of the CN feature: 7800--8200~\AA. This slope is defined to be the median of the last 25 flux values in this wavelength interval minus the median of the first 25 flux values, divided by length of the wavelength interval: 400~\AA. A linear ramp is added to the science spectrum in order to match its slope to the slope of the coadded spectral template and to ensure that the $\chi^2$ statistic (Eqn.~\ref{eqn:chi2}) is sensitive to the details of the spectral absorption lines and insensitive to the continuum slope.

With this final step completed, we are able to quantify the level of similarity each star in our sample has with both the C star and weak CN star templates by following what is effectively a modification of the $\chi^2$ formula:

\begin{equation}\label{eqn:chi2}
    \chi^2 = \frac{\sum_\lambda{\left[ f(\lambda) - f_{\rm templ}(\lambda) \right]^2 {\rm ivar}(\lambda)}}{\sum_\lambda{{\rm ivar}(\lambda)}}
\end{equation}
\noindent

\noindent
where $f$ and $f_{\rm templ}$ are the normalized flux values of the science spectrum (after slope matching) and coadded template spectrum (for weak CN or C stars), respectively, and the sum is computed over the 7800--8200~\AA\ wavelength interval.
We use this modified $\chi^2$ statistic to then ``score'' each star in our sample against both the weak CN and C star coadded spectral templates.

It is important to note that, when scoring M31 stars, we only compare them against spectral templates generated from M31 stars (and the same for M33 stars). We experimented with using coadded spectral templates generated from M31 stars to compare against the spectra of individual stars in M33 and vice versa, and found that it did not affect our COMET plot results significantly. Such a cross-comparison across the two galaxies led to a slight increase in the scatter of the data points in the resulting COMET plots (more on that in \S\,\ref{sec:m31m33comp} below). Regardless, for the remainder of this work, we have chosen to perform the analysis separately for the two galaxies.

\begin{figure}[H]
    \centering
    \includegraphics[width=0.7\textwidth]{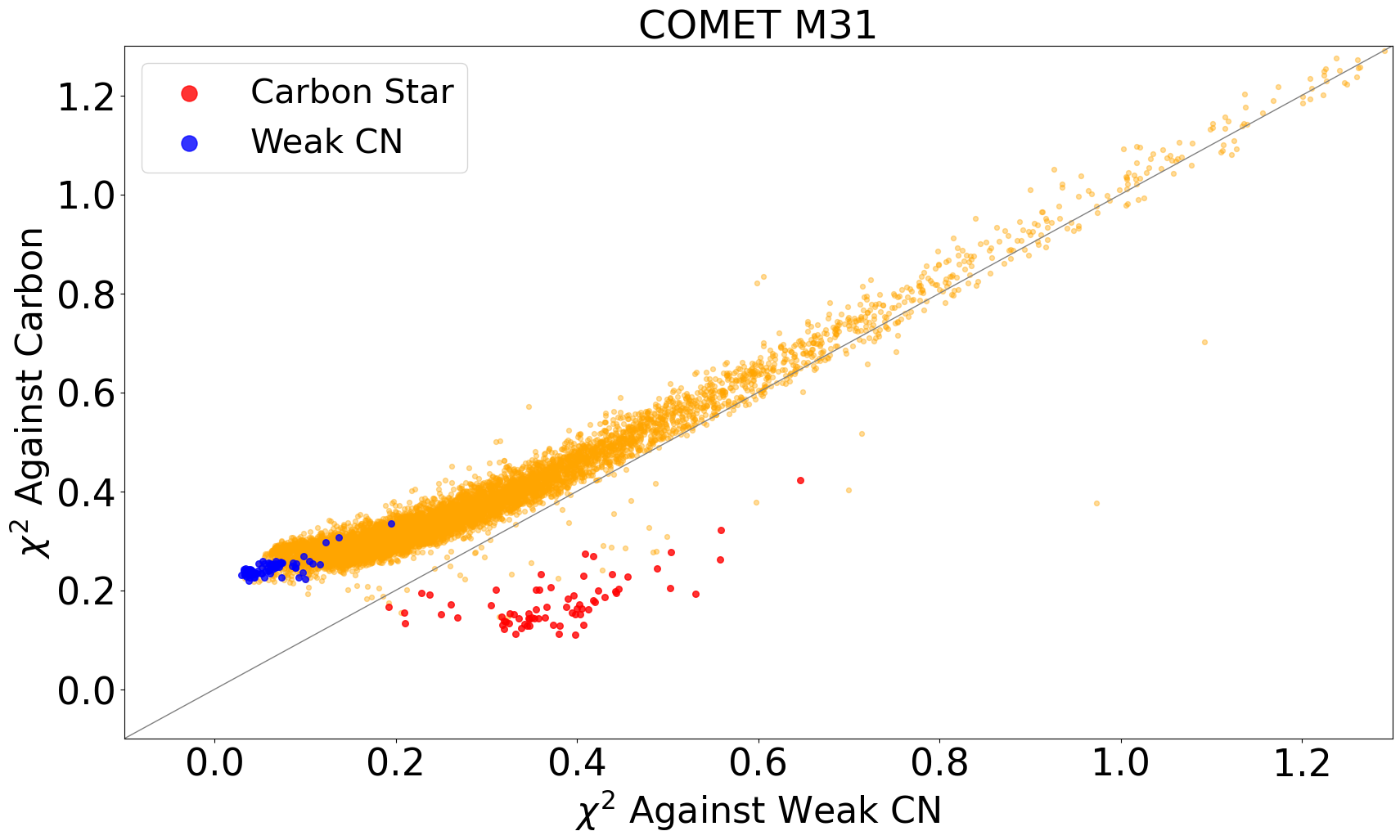}
    \caption{The COMET plot for M31 stars in which the modified $\chi^2$ score of each star relative to the weak CN star spectral template (x axis) is plotted against its score relative to the C star spectral template (y axis). Visually classified weak CN and C stars are plotted in blue and red, respectively. The remaining stars are plotted in orange. A 1:1 line is plotted in grey to separate the stars that are more similar to the weak CN star spectral template than the C star spectral template (above the grey line) and vice versa (below the grey line). Most of the sample, with the exception of C stars, lies above the 1:1 line, with the weak CN population at the head of a comet-like distribution.}
    \label{fig:comet1}
\end{figure}

In Figure \ref{fig:comet1}, we plot the modified $\chi^2$ score of each M31 star relative to the weak CN star coadded spectral template against its score relative to the C star coadded spectral template. We have named this simple strategy for differentiating between the weak CN and C populations of stars the \textit{Comparison to Empirical Templates} (COMET) method. This acronym (or, more accurately, ``backronym'') refers to the comet-like appearance of the distribution of stars, with weak CN and `normal' O-rich stars forming the comet head and tail, respectively, while the C stars form a distinct group in the lower-right part of the plot. Stars that lie above the diagonal 1:1 grey line are more similar to weak CN stars than to C stars, and vice versa for those below the line. Stars that lie near the head of the COMET plot can be thought of as weak CN stars. 

\begin{figure}[H]
    \centering
    \includegraphics[width=0.7\textwidth]{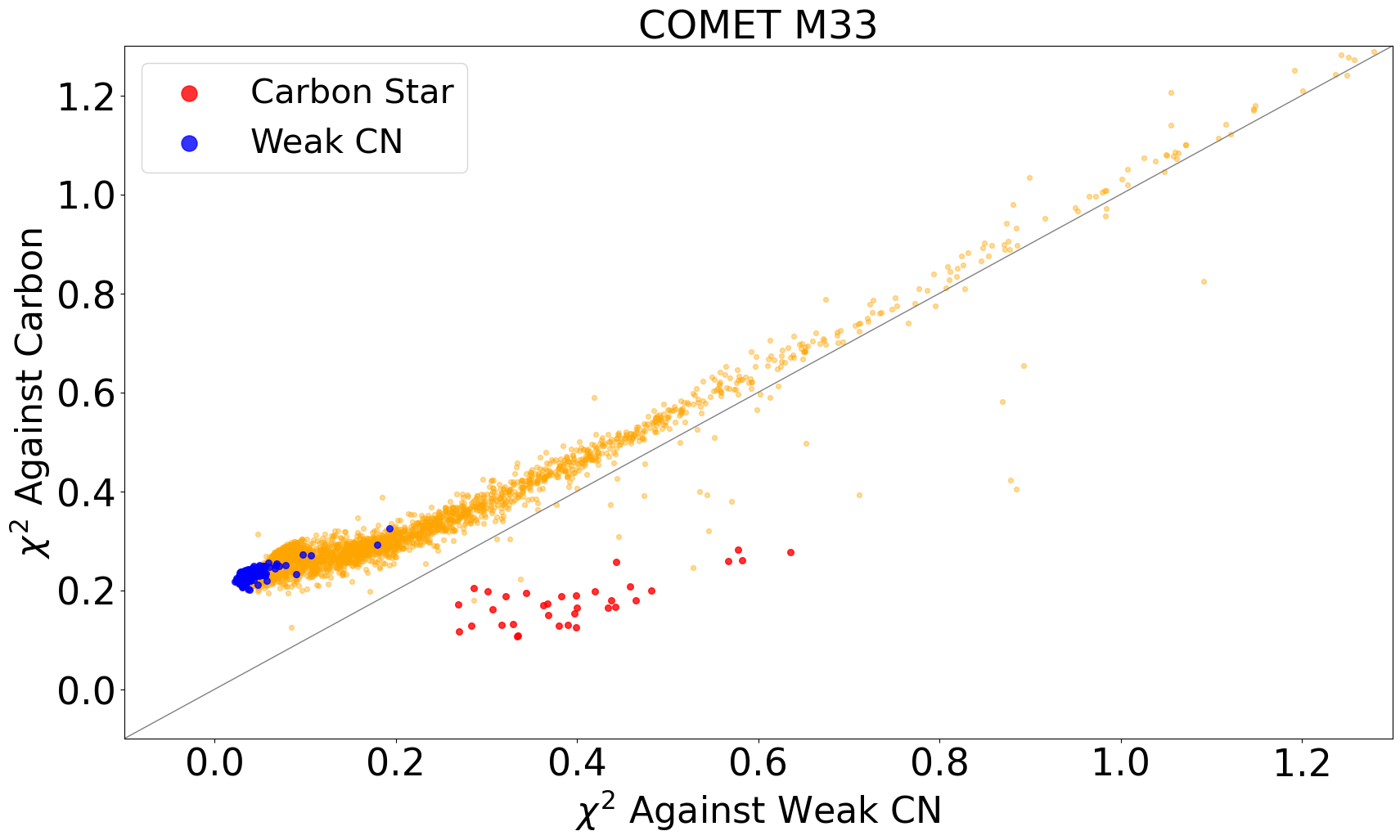}
    \caption{Same as Figure \ref{fig:comet1} but for M33 stars.}
    \label{fig:comet2}
\end{figure}

The COMET method is run on both M31 and M33 datasets (Figures \ref{fig:comet1} and \ref{fig:comet2}). The  sample of visually inspected weak CN and C stars are shown in blue and red, respectively. The rest of the stars are plotted in orange. The 1:1 line in these figures cleanly separates the C star population from the rest of the dataset. The distribution of data points in the COMET plot appears to indicate that the weak CN star population is {\it not\/} an evolutionary `cousin' of the C star population, as we had originally imagined. Instead, the weak CN population should be interpreted as more of a tail to the distribution of `normal' O-rich stars.

\subsection{Weak CN Parameter $\omega(d)$}\label{sec:omega}
As can be seen from Figures \ref{fig:comet1} and \ref{fig:comet2}, visually identified weak CN stars tend to lie near the head of the COMET and have a low $\chi^2$ (good fit) relative to the weak CN template, but the COMET head is well above the 1:1 line. This is only to be expected: the weak CN spectral template was constructed by coadding the spectra of a subset of the visually classified weak CN stars. The main difference between the spectra of visually classified weak CN stars and visually classified C stars in the 7800--8200\,\AA\ region is the strength of their CN feature. Therefore, it is no surprise that visually identified weak CN stars have a relatively low $\chi^2$ with respect to the C star spectral template. Indeed, of the entire sample above the 1:1 line, weak CN stars have the lowest $\chi^2$ with respect to the C star template. 

In order to objectively select weak CN stars, we define an exponential parameter:

\begin{equation}
    \omega(d) = \exp{(-d)}
    \label{eq:omega}
\end{equation}

\noindent where $d$ is the distance of each star in the COMET plot from the point (0.0, 0.2), which we arbitrarily define to be the origin.

\begin{figure}[H]
    \centering
    \includegraphics[width=0.99\textwidth]{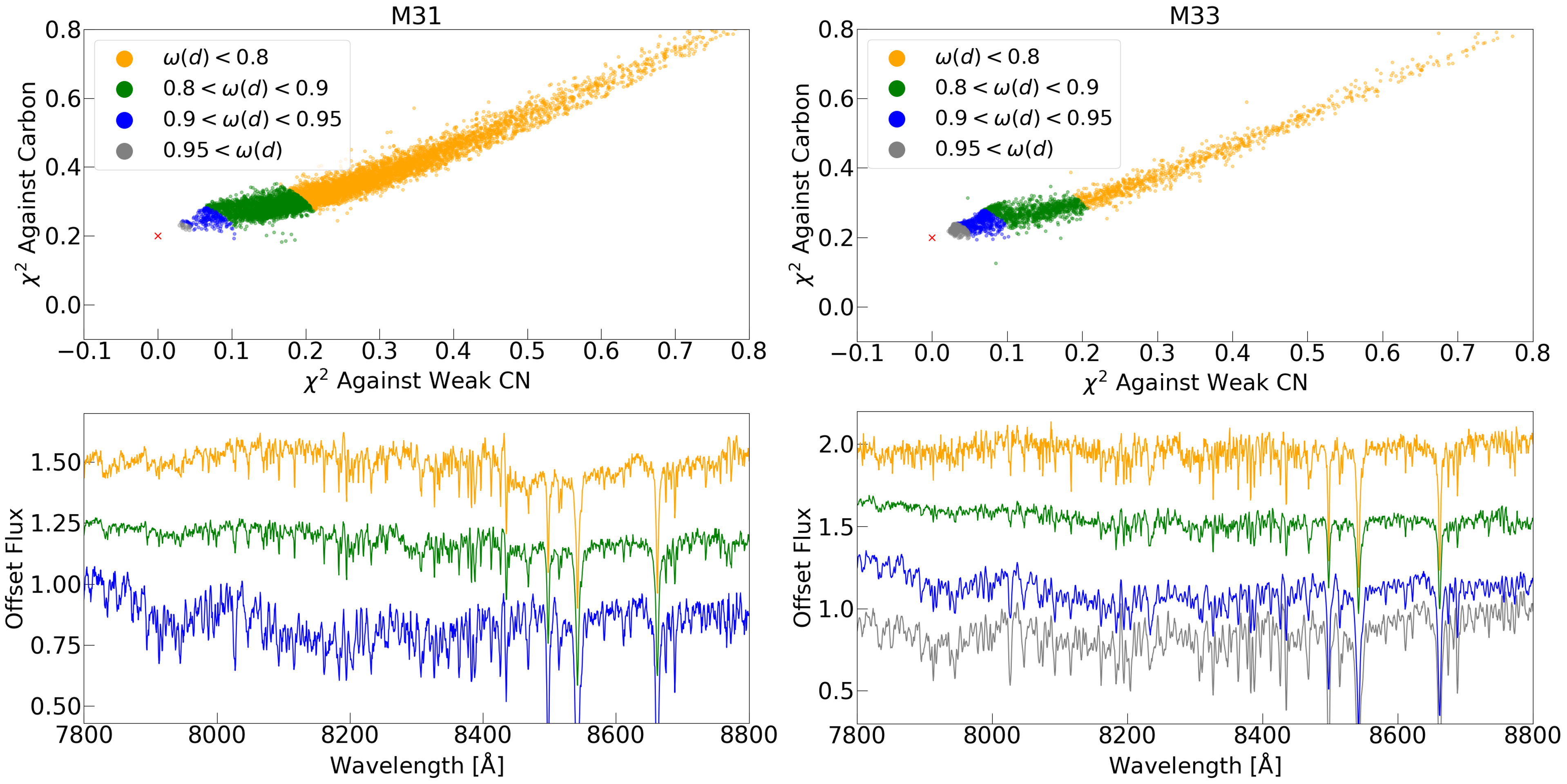}
    \caption{\textit{Top}: COMET plots for M31 (left) and M33 (right) in which stars are color coded according to their distance from the COMET head in four bins of $\omega(d)$. The small red `x' marks the origin relative to which the distance $d$ is measured. \textit{Bottom}: Coadded spectra as a function of distance from the COMET head for M31 (left) and M33 (right) for four subgroups of stars binned according to their $\omega(d)$ parameter; there are too few stars with $\omega(d)>0.95$ in M31 to create a coadded spectrum which explains why there is no gray curve in the lower left panel (see \S\,\ref{sec:m31m33comp} for a possible explanation). Arbitrary vertical offsets have been applied to the coadded normalized spectra for ease of viewing.}
    \label{fig:coadd1}
\end{figure}

To examine the relationship between the $\omega(d)$ parameter and the strength of the weak CN spectral absorption feature, we group stars into four bins: $\omega(d)<0.8$, $0.8<\omega(d)<0.9$, $0.9<\omega(d)<0.95$, and $\omega(d)>0.95$; this grouping is done for each of M31 and M33. Figure \ref{fig:coadd1} shows these subgroups of stars in M31 and M33 in the COMET plot and their corresponding coadded spectra (these coadds were generated in the same way as the weak CN and C star spectral templates in \S\,\ref{sec:templates}). Only stars above the 1:1 line in the COMET plot are included in the spectral coaddition. We do this to explicitly exclude C stars and limit the spectral coadd to weak CN and normal stars.

Figure~\ref{fig:coadd1} shows that the stars closest to the COMET's head show a clear weak CN absorption feature in their coadded spectrum (blue line in the lower left panel for M31; grey and blue lines in the lower right panel for M33). The COMET head is further from the origin for M31 stars than for M33 stars (upper panels); there are not enough M31 stars in the $\omega(d)>0.95$ bin for us to construct a coadded spectrum (more on that below and in \S\,\ref{sec:m31m33comp}). For both galaxies, the CN spectral feature is very weak or absent for subsequent bins along the COMET body/tail (green and orange curves), signifying that the vast majority of weak CN stars lie within the $\omega(d)>0.9$ bins.

\begin{figure}[H]
    \centering
    \includegraphics[width=0.99\textwidth]{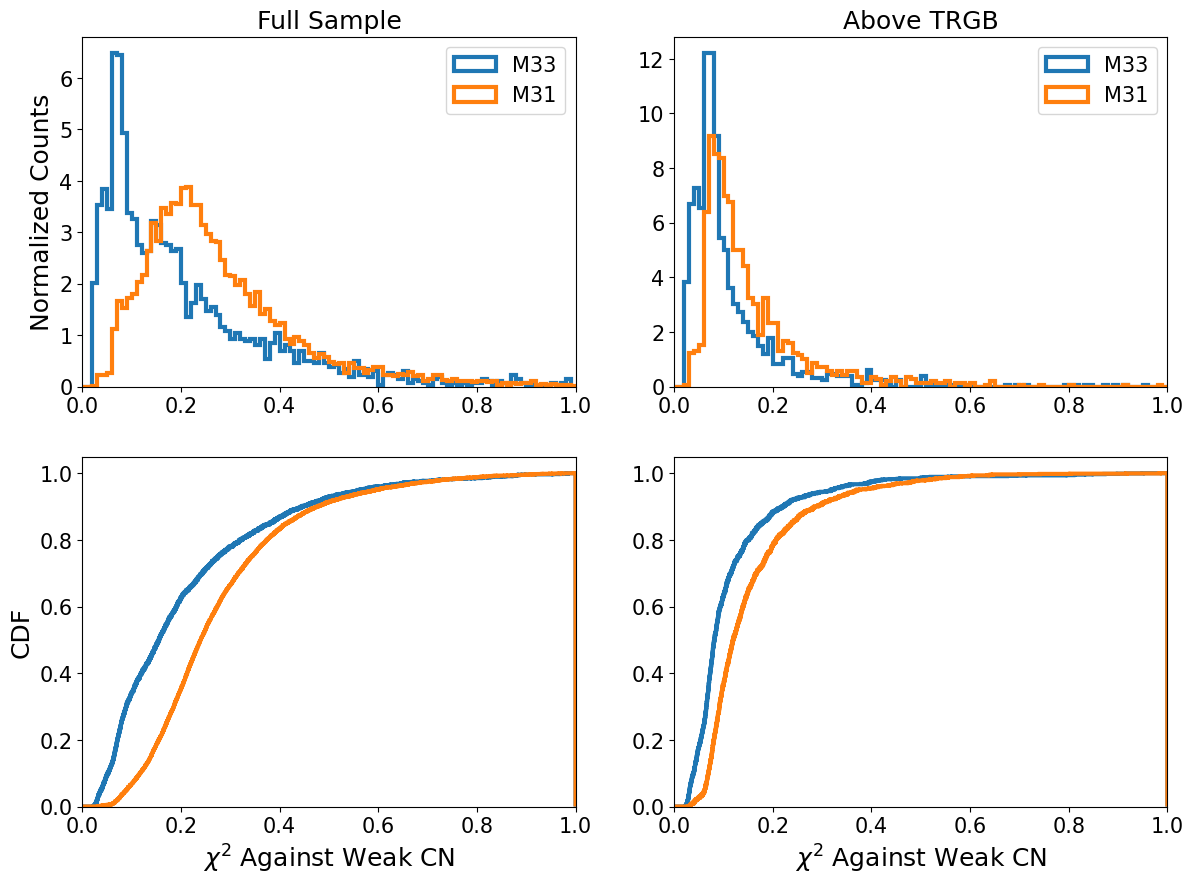}
    \caption{Normalized differential (top) and cumulative (bottom) histograms of modified $chi^2$ scores relative to the weak CN spectral template for M31 and M33 stars. The left panels are for our full M31 and M33 samples, while the right panels are for the super-TRGB subsamples. The full M33 sample is skewed towards smaller scores than the full M31 sample; the difference between the two galaxies, while smaller than for the full sample, is still clear for the super-TRGB samples (see \S\,\ref{sec:m31m33comp} for a possible explanation of the difference between M31 and M33).}
    \label{fig:cdf1}
\end{figure}

The M31 and M33 samples have different distributions in the COMET plot. The left panels of Figure~\ref{fig:cdf1} show the normalized differential (top) and cumulative (bottom) histograms of modified $\chi^2$ scores relative to the weak CN spectral template, and reveal a clear difference between the two galaxies in the sense that, on average, M31 stars have higher scores (i.e., are farther away from the COMET head) than M33 stars.

Before any astrophysical conclusions can be drawn, however, it is important to account/correct for differences in spectroscopic target selection between the two galaxies. The M31 sample goes 1--2~mag deeper in F814W than the M33 sample, as can be seen in the CMDs in middle columns of Figures~\ref{fig:CMD1} and \ref{fig:CMD2}. We apply an $\rm F814W<20.7$ cut to both M31 and M33 samples in order to limit them to massive evolved stars above the tip of the RGB (TRGB). This essentially eliminates the difference in the M31 vs.\ M33 spectroscopic target selection functions, while preserving all of the weak CN and C stars. The right panels of Figure~\ref{fig:cdf1}, like the left panels, show histograms of scores relative to the weak CN spectral template, but for only super-TRGB stars in M31 and M33. The difference between the two galaxies is smaller when one restricts the comparison to the super-TRGB subsamples, but M31 stars still have higher scores on average than M33 stars.

\begin{figure}[H]
    \centering
    \includegraphics[width=0.99\textwidth]{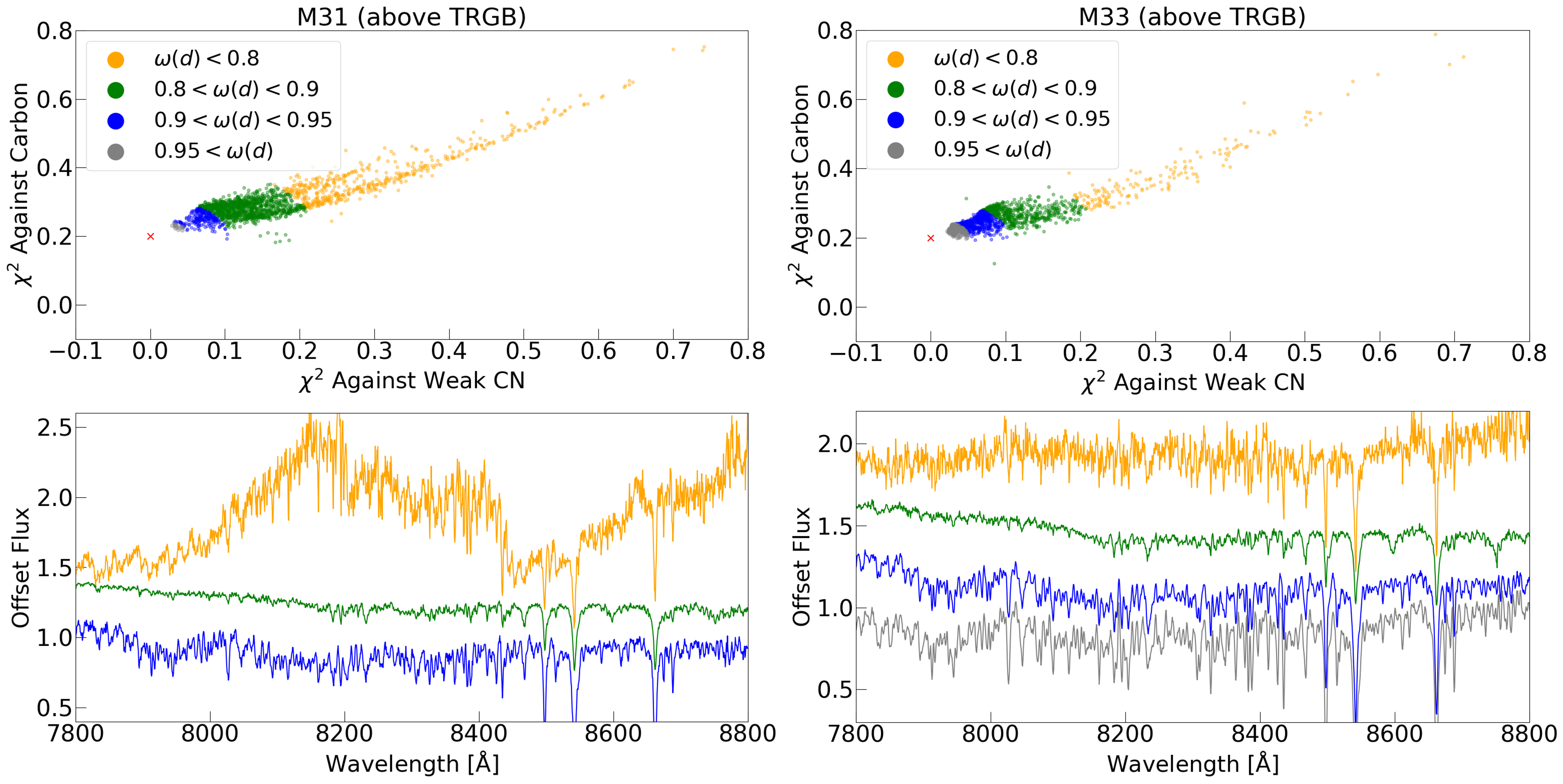}
    \caption{Same as Figure \ref{fig:coadd1}, but only for stars above the TRGB. The trends in the COMET plot and coadded spectra are the same as for the full samples and similar between M31 and M33, with the exception of stars with $\omega(d)<0.8$ (orange data points and spectra). These trends and differences are discussed in \S\,\ref{sec:m31m33comp}.}
    \label{fig:coadd3}
\end{figure}

Figure~\ref{fig:coadd3} shows the same combination of COMET plots and coadded spectra of $\omega(d)$-based subgroups as Figure~\ref{fig:coadd1}, except it is based on the super-TRGB subsamples of M31 and M33 stars. Limiting the samples to stars above the TRGB greatly reduces the length/prominence of the COMET tail. Based on the visibility/clarity of the weak CN spectral feature in the coadded spectra of the various subgroups shown in Figures~\ref{fig:coadd1} and \ref{fig:coadd3}, an $\omega(d)>0.9$ criterion appears reasonable for automated classification of weak CN stars. This criterion yields a sample of 224 weak CN stars for M31 and 659 weak CN stars for M33.

\section{The Weak CN Phenomenon in Red Supergiants}\label{sec:rsg}

\subsection{Photometric Properties of Weak CN Stars in Color-Magnitude Space}\label{sec:cmd}
\begin{figure}[H]
    \centering
    \includegraphics[width=0.99\textwidth]{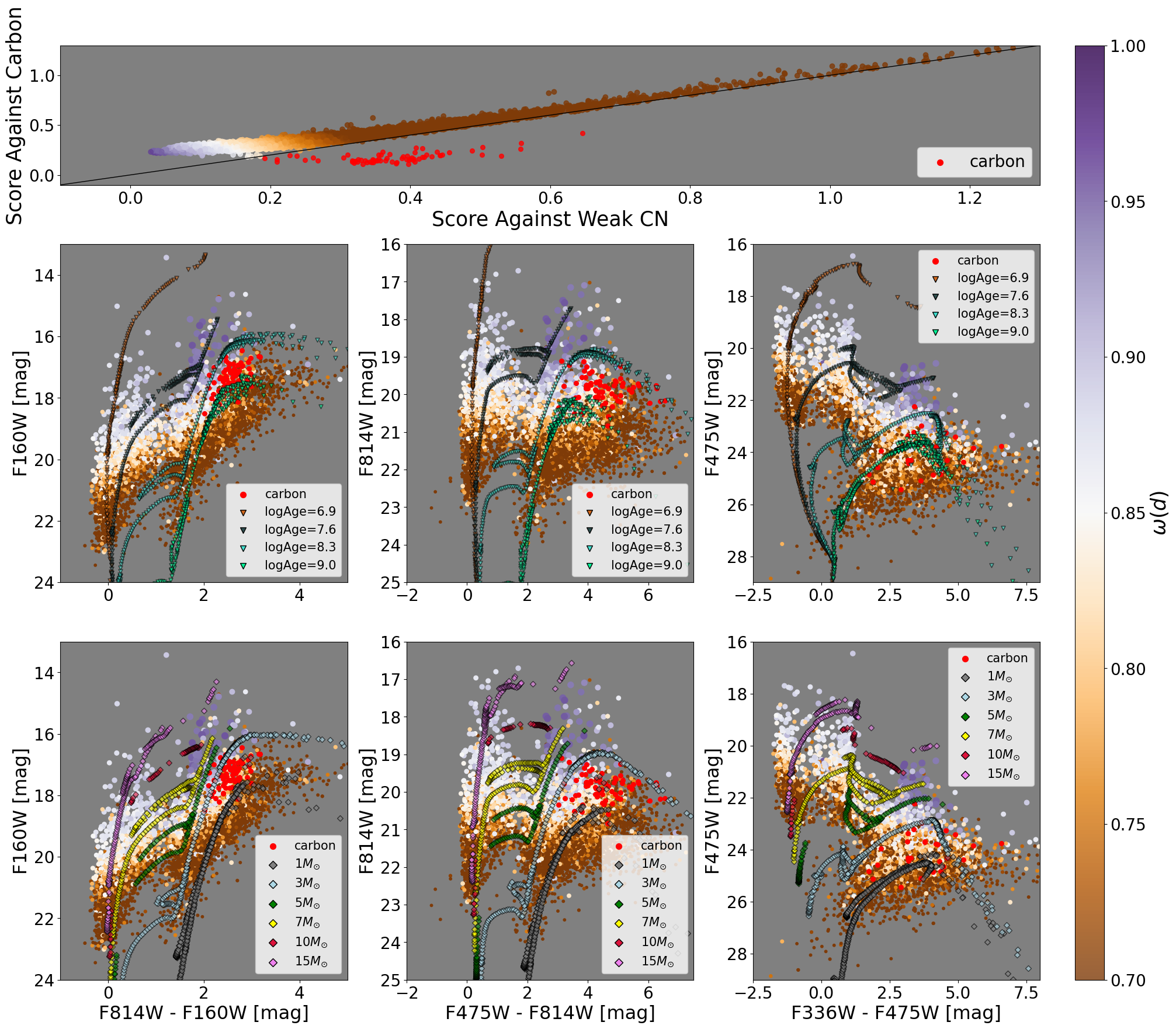}
    \caption{\textit{Top}: COMET plot for M31 stars, in which all but the C stars are colored by the parameter $\omega(d)$, which is an exponential function of the distance of the star from the COMET head (\S\,\ref{sec:omega}); C stars are marked as red dots. \textit{Middle Row}: CMDs showing the position of weak CN and other stars in M31 in various {\it HST\/} filter color-magnitude combinations, using the same color coding scheme for the stars as the COMET plot in the top panel. \textit{Middle Row-Left Column}: $H$ vs.\ $I-H$. \textit{Middle Row-Center Column}: $I$ vs.\ $B-I$. \textit{Middle Row-Right Column}: $B$ vs.\ $U-B$. Isochrones (loci of constant stellar age) are overlaid on each CMD panel; they show that the weak CN population corresponds to a fairly short-lived population of stars with lifetimes of a few tens of Myr. \textit{Bottom Row}: Same CMD data as the middle row of panels, but with tracks of constant stellar mass overlaid. In M31, the weak CN population ranges masses of 3--$10\, M_{\odot}$.}
    \label{fig:CMD1}
\end{figure}

\begin{figure}[H]
    \centering
    \includegraphics[width=0.99\textwidth]{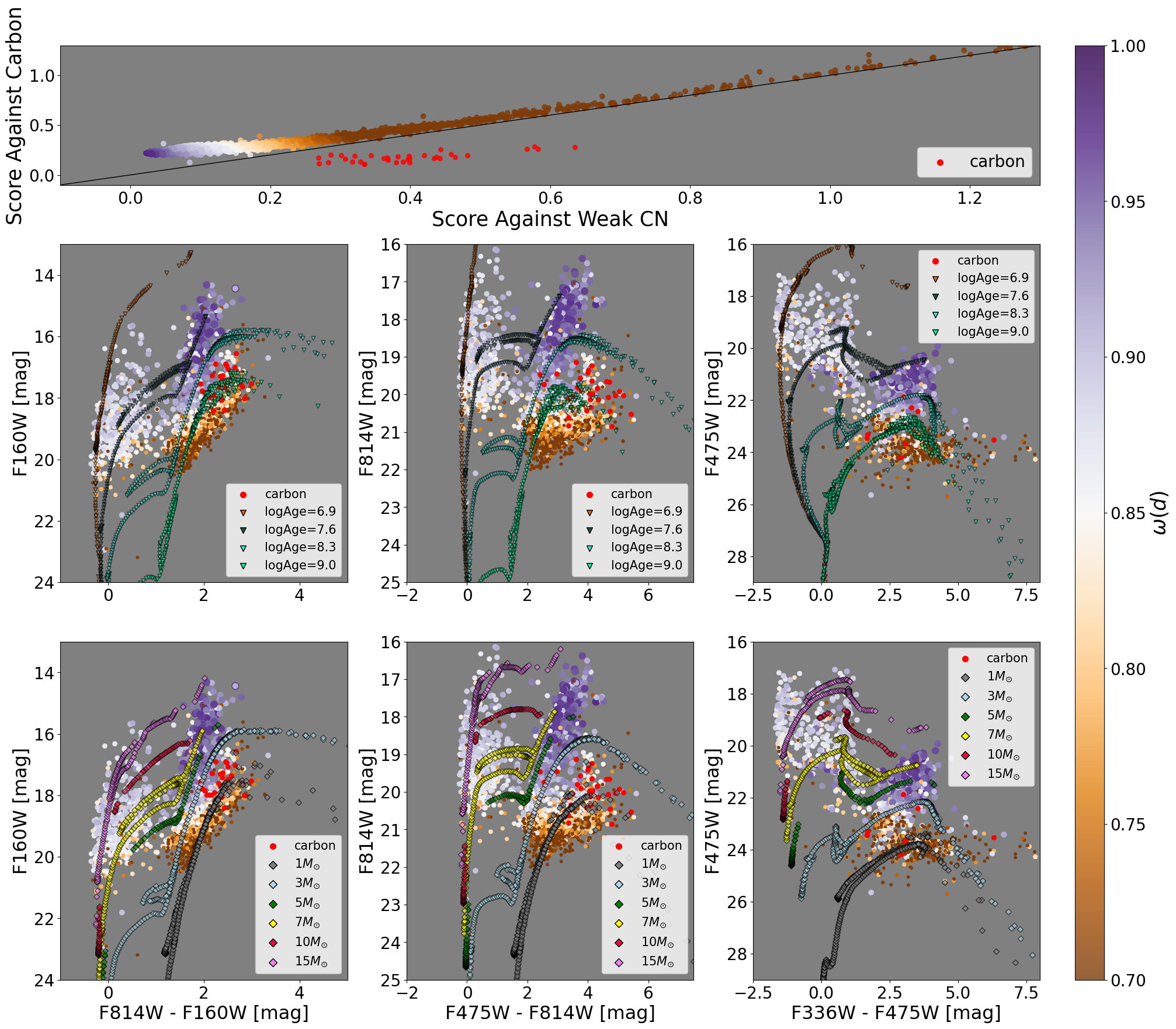}
    \caption{Same as Figure \ref{fig:CMD1} but for M33. We note that the M33 sample does not contain as many faint stars or red stars as the M31 sample (this is clearest in the optical and near infrared CMDs in the center and left column of panels, respectively). The former difference is the result of a difference in spectroscopic target selection criteria between our M31 and M33 Keck DEIMOS surveys, while the latter difference is the result of M31 having a higher mean metallicity. We also note the much larger fraction of stars with large $\omega(d)$ values in M33. The estimated age and mass ranges of weak CN stars in M33 matches those of their M31 counterparts.}
    \label{fig:CMD2}
\end{figure}

Figures~\ref{fig:CMD1} and \ref{fig:CMD2} show the distribution of weak CN, C, and other stars in the CMD for M31 and M33, respectively. C stars are marked as red dots, while the rest of the stars are colored based on the exponential distance parameter $\omega(d)$ --- i.e., how similar each star's spectrum is to the weak CN spectral template --- such that higher $\omega (d)$ values correspond to lower $\chi^2$ values. It is possible to see a general trend in which more luminous, or massive, stars are more alike to weak CN stars. This could be a simple correlation between luminous stars having a higher signal-to-noise than fainter stars, and thus on average matching our high signal-to-noise templates better. Regardless, the group of stars colored in shades of purple, which comprises the stars that are close to the head of the COMET distribution and, therefore, are most like weak CN stars, is very well localized to the red core He-burning sequence. In the companion paper to this work, \citet{GrionFilho25} find that weak CN stars have an average age of $5.6\times10^7$~yr for M31 and $4.2\times10^7$~yr for M33. Based on Figures~\ref{fig:CMD1} and \ref{fig:CMD2}, weak CN stars appear to be short-lived massive red supergiants. This conclusion is firmed up in the next subsection, where we present constant-age model isochrones and constant mass model stellar tracks overlaid on the CMDs. 

\subsection{Comparison to Model Isochrones and Stellar Tracks}\label{sec:models}
For the analysis in this subsection, we use the PARSEC set of model isochrones and tracks \citep{Bressan12} to cover all phases of stellar evolution from the pre-main sequence up to C ignition in massive stars, or up to the occurrence of the first thermal pulse in low- and intermediate-mass stars. In order to cover our C star sample, the TP-AGB phase is then computed with the COLIBRI code \citep{Marigo13,Pastorelli19, Pastorelli20}. These model isochrones and tracks are calculated in the {\it HST\/} filter system and shifted by the apparent distance modulus and reddening of M31 and M33. Based on extinction coefficients in this filter system that are applicable for cool stars \citep[see Fig.~2 of][]{Girardi08}, the apparent distance modulus in F814W is $(m-M)_{\rm F814W}=25.0$ and 24.9~mag for M31 and M33, respectively, and the reddening is $E({\rm F475W-F814W})=0.6$ and 0.2~mag for M31 and M33, respectively.

In the upper row of CMDs in Figures \ref{fig:CMD1} and \ref{fig:CMD2} (middle row of panels), the red He-burning feature of the isochrone with an age of $t=40~\rm Myr = 10^{7.6}~yr$ appears to be the best match to the location of weak CN stars, with C stars being a better match to significantly older, intermediate-age isochrones \citep{Mouhcine02}. The large number of weak CN stars (224 for M31 and 659 for M33) indicates that it is a relatively long-lived phenomenon, making them unlikely candidates for AGB stars. Indeed, lifetimes on the red part of the core He-burning sequence are typically a few Myr which is substantially longer than the sub-Myr lifetimes of AGB stars with hot bottom burning and super-AGB stars \citep{Bressan12}. This is consistent with our initial assessment that weak CN stars are a group of relatively short-lived evolved massive stars. Given their clustered position in the CMD and these initial estimates for age and mass, we believe that the weak CN feature must be a common stage in the evolution of short-lived massive red supergiants, which has been overlooked so far.

In the bottom row of CMD panels in Figures \ref{fig:CMD1} and \ref{fig:CMD2}, model tracks with stellar masses $3 M_{\odot} \leq M_{\rm *} \leq 10 M_{\odot}$ correspond to the region in the CMD where weak CN stars lie, with most of the group lying close to the $7 M_{\odot}$ track. By contrast, C stars lie in a region of the CMD that corresponds to tracks with substantially lower stellar masses, ranging from 1--3 $M_{\odot}$.

\section{Discussion}\label{sec:disc}

\subsection{Comparison of Weak CN stars in M31 and M33}\label{sec:m31m33comp}

In this subsection, we examine the similarities and differences between M31 and M33 in terms of various properties of their weak CN star and other stellar populations: their coadded spectra, COMET plot distribution, and CMD distribution. Where applicable, we offer possible astrophysical explanations of these similarities and differences.

\begin{itemize}

\item Let us start with some of the \textit{similarities} displayed by the two galaxies. With the exception of C stars, which are excluded from the coadded spectra of $\omega(d)$-based subgroups shown in Figure~\ref{fig:coadd3}, all other subgroups of stars display the near infrared Ca~II triplet of absorption lines ($\lambda\lambda 8498, 8542$, and 8662\,\AA) that is characteristic of O-rich stars. Moreover, the coadded spectra for the $0.9<\omega(d)<0.95$ and $0.8<\omega(d)<0.9$ subgroups look very similar between M31 and M33. For the higher $\omega(d)$ subgroup (blue lines in the lower panels), the weak CN spectral feature is obvious and similar between the two galaxies---this includes the subtle ``W'' shape in the 7800--8200\,\AA\ range with the broad bump at 8000\,\AA\ and the details of the finer spectral features. For the lower $\omega(d)$ subgroup (green lines in the lower panels), the weak CN spectral feature is comparably weak/marginal in the two galaxies. For this $0.8<\omega(d)<0.9$ subgroup in M31 and M33, the spectral feature is too weak/subtle for it to be discernible in \textit{individual} spectra given their typical signal-to-noise ratio (e.g.,~see Figures~\ref{fig:spectracomp1} and \ref{fig:spectracomp2}). However, the fact that the stars in this subgroup span a range of $\omega$ values, corresponding to a range of modified $\chi^2$ scores relative to the weak CN template of $\approx0.1$--0.2, indicates that their spectra span a range of weak CN strengths.


\item The first and most obvious \textit{difference} between the two galaxies is in their weak CN star fractions: 12\% in M31 (224 weak CN stars out of a sample of 1889 super-TRGB stars) versus 47\% in M33 (659 weak CN stars out of a sample of 1406 super-TRGB stars). The red supergiant sequence that the weak CN phenomenon appears to be associated with---e.g.,~stars in the color range $\rm F475W-F814W=2$--4 and magnitude range $\rm F814W=17$--20 in the center column of Figures~\ref{fig:CMD1} and \ref{fig:CMD2}---is much more prominent in M33 than M31. This is because M33's specific star formation rate (SSFR), defined to be the star formation rate per unit stellar mass, is about an order of magnitude higher than M31's. For example, recent studies using PHAT and PHATTER data have shown that M31 and M33 have comparable star formation rates \citep{Lewis15, Lazzarini22}, while M33's stellar mass is $<10\%$ that of M31.

\item The second \textit{difference}, which has been mentioned earlier, is that, on average, M33's weak CN stars have lower modified $\chi^2$ scores with respect to the weak CN template than their M31 counterparts (this is illustrated by the relative paucity of grey data points in M31 in Figures~\ref{fig:coadd1} and \ref{fig:coadd3} and quantified in Figure~\ref{fig:cdf1}). In other words, the spectra of M33 weak CN stars are generally more similar to the M33 weak CN spectral template than those in M31 are to the M31 weak CN spectral template. This may be related to the fact that M33's red supergiant sequence displays a smaller spread in color than M31's. Two factors affect the color spread: (1)~M33 has a lower mean metallicity and smaller metallicity spread than M31 \citep{Rosolowsky08, Sanders12, Gilbert14}; and (2)~M33's disk is more face on than M31's so there are smaller reddening variations from star to star. Dust reddening should not affect the modified $\chi^2$ score since we match the spectral continuum slope of each star to that of the template. On the other hand, perhaps metallicity variations from star to star result in variations in their weak CN spectral feature. If the spectra of M33's weak CN stars are more homogeneous than those in M31, they will of course tend to be more similar to the spectral template, since the template is a coaddition (average) of those very same weak CN star spectra.


\item The third \textit{difference} between M31 and M33 is in the properties of the subgroup of super-TRGB stars with $\omega(d)<0.8$---i.e.,~those whose spectra are the most dissimilar to the weak CN template (orange data points and lines in Figure~\ref{fig:coadd3}). M31 stars display a clear bifurcation in the COMET tail, while no such bifurcation is seen in M33. The spectrum of this subgroup of stars shows TiO bands for M31, but not for M33. The lower portion of M31's bifurcated COMET tail is comprised of relatively blue (hot) stars and the upper portion is comprised of relatively red (cool) stars. The M31 super-TRGB sample extends to very red colors: $\rm F475W-F814W=0$--7, whereas that color range is 0--5 for M33 (compare Figures~\ref{fig:CMD1} and \ref{fig:CMD2}). This difference is related to the fact that the M31 super-TRGB sample, unlike the M33 super-TRGB sample, contains a substantial population of very high (even super-solar) metallicity AGB stars \citep{Rosolowsky08, Sanders12, Gilbert14}.

\end{itemize}


\subsection{Weak CN Stars in the Milky Way and Other Local Group/Local Volume Galaxies}\label{sec:mw_etc}

The weak CN spectral absorption feature appears to be an identifiable tracer of massive short-lived evolved stars, which tend to be found in areas of active star formation in their host galaxies. The connection with the core He-burning sequence and consequently red supergiants, as explained in \S\,\ref{sec:models} and Paper~II, offers tantalizing possibilities for the study of weak CN stars in the MW and other galaxies. From our vantage point within the disk of the MW, all but the nearest star forming regions are heavily obscured by interstellar dust, at least at optical wavelengths where this weak CN feature has been discovered. Well-studied MW red supergiants like Betelgeuse and Antares, which happen to be very close to us and are very bright, are excellent targets to search for and characterize the properties and possible time dependence of the weak CN spectral absorption feature (Grion~Filho et~al., in preparation). Red supergiants in low-mass star-forming galaxies in the Local Group/Local Volume, including, in particular, the Small Magellanic Cloud, could also serve as excellent targets for the same purpose.

As things stand, we are unsure about the exact physical mechanisms that would explain the prevalence of weak CN absorption among massive red supergiant stars. We present a more detailed analysis using evolutionary models in the companion paper to this work (Paper~II). It is our hope that, ultimately, high resolution, high signal-to-noise, time-resolved spectroscopy of MW red supergiants like Betelgeuse and Antares, along with large, high quality, spectroscopic samples of weak CN stars  across a variety of nearby host galaxies, will reveal interesting observational patterns that will provide insight into the astrophysical origin of this phenomenon.

\section{Conclusions and Future Work}\label{sec:conc}

%
%
%
%
%
%
%
The main points of this paper can be summarized as follows:

\begin{enumerate}

\item 
We have discovered that a subpopulation of stars in M31 and M33 have spectra that contain a weak CN absorption feature at 8000\,\AA, {$\sim 6$}$\times$ weaker than the corresponding CN absorption feature in C stars. For these weak CN stars, the rest of the spectrum resembles that of `normal' O-rich stars in that it contains absorption features like TiO bands and the Ca II near infrared triplet.

\item
While the weak CN spectral absorption feature is unexpected, it appears to occur in an otherwise well-understood subpopulation of stars. Based on the CMD location and theoretical isochrones and tracks, weak CN stars appear to be core He burning massive ($\approx5$--10\,$M_\odot$) red supergiants whose ages are of the order of 40--50~Myr.

\item
We generated a set of high signal-to-noise spectral templates from visually identified C star and weak CN star samples. We then scored each star in our overall spectroscopic sample according to their similarity to these templates in the COMET plot and found a strong clustering of the weak CN population near the head of the COMET. Based on the distance of each star from the head of the COMET, we identify a final sample of 224 weak CN stars in M31 (out of a full sample of 8009 stars and a super-TRGB subsample of 1889 stars) and 659 weak CN stars in M33 (out of a full sample of 2665 stars and a super-TRGB subsample of 1406 stars).

\item
In the particular projection of the spectroscopic data that the COMET plot represents, weak CN stars appear to occupy the tail of the distribution of ``normal'' O-rich stars, while C stars appear to be a disjoint population.

\item
The higher fraction of weak CN stars in M33 is consistent with the fact that this galaxy has a higher specific star formation rate than M31.

\end{enumerate}

Given that weak CN stars are short-lived red supergiants, we expect to find them only in regions where star formation is still active in the disks of their host galaxies. This could explain why the weak CN phenomenon has not been reported in the MW. Since most stars in the MW disk are heavily extincted/reddened by interstellar dust from our perspective, it is difficult (perhaps impossible) to obtain spectra at optical wavelengths (covering the 8000~\AA\ CN region) at high enough resolution to discern the weak CN feature. For very nearby red supergiant stars like Betelgeuse and Antares, however, the prospects are more optimistic, given that these stars are close enough to have low extinction due to interstellar dust (Grion~Filho et~al., in preparation).



While the weak CN and C star datasets presented in this paper are limited to Keck DEIMOS spectra collected during the 2021B and earlier observing seasons, we have continued to expand the SPLASH and TREX surveys during the 2022B, 2023B, and 2024B observing seasons. This expansion includes new DEIMOS masks in the PHAT region of M31, PHATTER region of M33, and the recent PHAST {\it HST\/} survey region which covers M32 and the southwestern disk of M31. Future papers will include these newer Keck DEIMOS spectra and present more complete samples of weak CN and C stars across the disks of M31 and M33.

\section*{Acknowledgments}

This paper is dedicated to the memory of our colleague Paola Marigo. We thank Anil Seth, Katie Hamren, Emily Cunningham, Tiburon Batriedo, Allison Chang, Swathya Chauhan, Sumedh Guha, Jon Hays, Suhas Kotha, Luiza Montecchiari, Atmika Sarukkai, Alyssa Sales, and Dhruv Trivedi for their contributions to the early stages of this project. We thank Emily Levesque and Bruce Margon for useful discussions. AB, AK, AlM, and ArM participated in this research under the auspices of the Science Internship Program (SIP) at the University of California Santa Cruz. We acknowledge support from National Science Foundation grants AST-2206328 and AST-1909759 and NASA/STScI grants HST-GO-14610, HST-GO-12055, HST-GO-14072, HST-GO-15932, and HST-GO-16274.

\bibliographystyle{aasjournal}
\bibliography{paper_rev4.bbl}

\end{document}